\documentclass[conference]{IEEEtran}
\usepackage{graphicx}
\usepackage{setspace}
\usepackage{amsmath}
\usepackage{amssymb}
\usepackage{bm}
\usepackage{cite}
\usepackage{float}
\usepackage[usenames,dvipsnames]{pstricks}
\usepackage{epsfig}
\usepackage{pst-grad} % For gradients
\usepackage{pst-plot} % For axes
\usepackage{array}
\usepackage[keeplastbox]{flushend}
\allowdisplaybreaks
\usepackage[letterpaper, left=0.625in,right=0.625in,top=0.75in,bottom=1.042in]{geometry}

\newcommand{\nn} {\nonumber}

\begin{document}

\newcommand{\beq}{\begin{equation}}
\newcommand{\beqarr}{\begin{eqnarray}}
\newcommand{\beqarrn}{\begin{eqnarray*}}
\newcommand{\eeq}{\end{equation}}
\newcommand{\eeqarr}{\end{eqnarray}}
\newcommand{\eeqarrn}{\end{eqnarray*}}

\def\undb#1{\mbox{\bf{#1}}}

\title{Incremental Selective Decode-and-Forward Relaying for Power Line Communication}

\author{
    \IEEEauthorblockN{Ankit~Dubey\IEEEauthorrefmark{1}, Chinmoy~Kundu\IEEEauthorrefmark{2}, Telex~M.~N. Ngatched \IEEEauthorrefmark{3}, Octavia A. Dobre \IEEEauthorrefmark{3}, and Ranjan K. Mallik \IEEEauthorrefmark{4}}
    \IEEEauthorblockA{\IEEEauthorrefmark{1}Department of ECE, National Institute of Technology Goa, Farmagudi, Ponda, Goa 403401, India}
    \IEEEauthorblockA{\IEEEauthorrefmark{2}School of Electronics, Electrical Engineering and Computer Science, Queen's University Belfast, U.K.}
     \IEEEauthorblockA{\IEEEauthorrefmark{3}Department of ECE, Faculty of Engineering and Applied Science, Memorial University, Canada}
    
    \IEEEauthorblockA{\IEEEauthorrefmark{4}Department of Electrical Engineering, Indian Institute of Technology Delhi, Hauz Khas, New Delhi 110016, India
    \\\textrm{\IEEEauthorrefmark{1}ankit.dubey@nitgoa.ac.in}, {\IEEEauthorrefmark{2}c.kundu@qub.ac.uk},{\IEEEauthorrefmark{3} tngatched@grenfell.mun.ca, odobre@mun.ca}, \IEEEauthorrefmark{4}{rkmallik@ee.iitd.ernet.in}}
    }

\maketitle

\begin{abstract}
In this paper, an incremental selective decode-and-forward
(ISDF) relay strategy is proposed for power line communication (PLC) systems to improve the spectral efficiency. Traditional decode-and-forward (DF) relaying employs two time slots by using half-duplex 
relays which significantly reduces the spectral efficiency. 
The ISDF strategy utilizes the relay only if the direct link
quality fails to attain a certain information rate, thereby improving the spectral efficiency.
The path gain is assumed to be log-normally distributed with very high distance dependent 
signal attenuation. Furthermore, the additive noise is modeled as a Bernoulli-Gaussian 
process to incorporate the effects of impulsive noise contents. 
Closed-form expressions for the outage probability and the fraction of times
the relay is in use, and an approximate closed-form expression
for the average bit error rate (BER) are derived for the  binary phase-shift keying signaling scheme.
We observe that the fraction of times the relay is in use can be significantly reduced compared to the traditional DF strategy.
It is also observed that at high transmit power, the spectral efficiency
increases while the average BER decreases with increase in the required rate. 
%Interestingly, it is also observed that while at a higher transmit power the spectral efficiency increases, the average BER decreases when the required rate increases. 
\end{abstract}


\IEEEpeerreviewmaketitle

\section{Introduction}
Various modern concepts, such as home automation, real-time energy monitoring system, and smart grid rely primarily on communication systems. Furthermore, the power-line communication (PLC) is a 
key solution by providing greater appliance-to-appliance connectivity \cite{PLC_BOOK:10, ShMaMiDuRa:16}. 

Although PLC is one of the preferred communication solutions for such applications,
it has various challenges.
The communication signals transmitted through power lines suffer from additive distortion, which comprises both background and impulsive noise \cite{PLC_BOOK:10}. In addition to additive distortions, communication signals are also affected by multiplicative distortions. Cables used to carry high amplitude alternating power signals at very low frequency (around $50$ or $60$ Hz) become hostile when carrying low amplitude communication signals at very high frequency; hence, communication signals undergo a heavy distance-dependent attenuation \cite{PLC_BOOK:10}. Furthermore, due to reflections from various terminations, multi-path propagation occurs and causes the received signal strength to fluctuate with time.
In most cases, the envelope of these fluctuations follows the log-normal distribution \cite{PaCaKaTh:03, GuCeAr:11, Ga:11}.
Thus, for reliable long-distance communication, it is essential to mitigate the effects of additive and multiplicative distortions.
A well established technique of relay-based communication is therefore proposed for PLC \cite{LaScYi:06, LaVi:11}. For multi-hop transmission, a distributed space-time coding technique is introduced
in \cite{LaScYi:06}, while a cooperative coding for narrowband PLC is proposed in \cite{LaVi:11}.
%In \cite{YoJaKiBa:14}, an opportunistic routing for PLC access networks is presented.
The average bit error rate (BER) and outage probability analysis using decode-and-forward (DF) relay is studied in \cite{DuMaSc:15}, omitting the direct transmission. Recently, in \cite{DuMa:16}, the correlation among multi-hop channels has also been considered for closely-placed DF relays; still, the direct transmission is ignored. Further, very recently, a class of machine learning schemes, namely multi-armed bandit, is proposed to solve the relay selection problem for dual-hop transmission in \cite{NiVi:17}.   

These works make use of half-duplex relays, which requires two time slots for the end-to-end communication. %\cite{Laneman_Wornell_cooperative_diversity}.
%\cite{MH_CR_BOOK:08, Laneman_Wornell_cooperative_diversity}.
The time slots required for the end-to-end communication can be significantly improved by using the incremental relaying strategy \cite{Laneman_Wornell_cooperative_diversity}, thereby improving the communication rate or spectral efficiency. In incremental relaying, the relay is used only if the direct transmission from source to destination fails to achieve a required information rate or equivalently a certain signal-to-noise (SNR) threshold. 

In conjunction with DF relays, incremental relaying can be applied with selective relaying called incremental selective DF (ISDF) relaying, whereby the relay is used only when the direct transmission fails and also the source to relay link achieves the required information rate. Though incremental or ISDF relaying has been investigated in wireless systems (see, e.g., \cite{Laneman_Wornell_cooperative_diversity,
%ikki_Performance_analysis_of_incremental,
Bai_Yuan_Performance_Analysis_of_SNR_Based}
and references therein), to the best of our knowledge, it has not been studied in PLC systems yet.

Motivated by the above discussions, in this paper, the ISDF strategy is proposed to enhance the spectral efficiency of PLC systems. The outage probability and average BER performance are evaluated. The PLC channels are assumed to follow the log-normal distribution with high distance-dependent attenuation, and the additive noise is assumed to follow the Bernoulli-Gaussian process. To get insight into the spectral efficiency, the fraction of times the relay is in use is also derived. Our main contributions are: i) to propose ISDF relaying for PLC systems to increase spectral efficiency, and ii) finding closed-form expressions for the outage probability and the fraction of times the relay is in use, and an approximate closed-form expression for the average BER considering the binary phase-shift keying (BPSK) signaling scheme.
%, in order to understand the system performance.

The rest of the paper is organized as follows. Section \ref{sec_system_model} describes the system model, while closed-form expressions for the outage probability and the fraction of times the relay is in use, and an approximate closed-form expression for the average BER are derived in Sections \ref{sec_outage}, \ref{sec_relay_usage}, and \ref{sec_aber}, respectively. Section \ref{sec_results} presents numerical and simulation results, and Section \ref{sec_conclusion} provides concluding remarks. 
%Acknowledgment is provided in Section \ref{sec_acknowledgment}. 

\textit{Notation:} $\mathbb{E}[\cdot]$ denote the expectation of its argument over the random variable (r.v.)
$X$, $\mbox{Pr}(\cdot)$ is the probability of an event, $P_e(\cdot)$ is the probability of bit error or BER,
$F_X (\cdot)$ represent the cumulative distribution function (CDF) of the r.v. $X$, and
$f_X (\cdot)$ is the corresponding probability density function (PDF).

\section{System Model}
\label{sec_system_model}

\begin{figure}
% [t]
\hspace{1.0cm}
\psfig{file=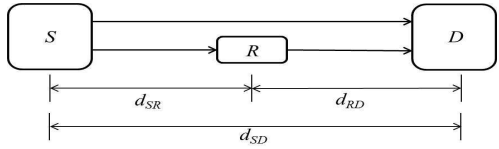,width=6.50cm,height=2cm}
\caption{PLC system model.}
\label{fig_1}
\end{figure}

The PLC system, as shown in Fig. \ref{fig_1}, consists of a source $S$, a destination $D$, 
and an ISDF relay, $R$.  $S$ tries to communicate with $D$ over a power cable with the help of $R$ 
to increase spectral efficiency. 
A link between any two nodes is denoted by $i\in\{SD, SR, RD \}$ where 
$SD$, $SR$, and $RD$ represent the links between $S$-$D$, $S$-$R$, and $R$-$D$, respectively. 
The length of the power cable between any two nodes is denoted by $d_i$, and $d_{SR}+d_{RD}=d_{SD}$.
In the first phase, $S$ broadcasts a symbol with power $P_{T_{S}}$. 
A predefined rate $R_{th}$ bits/sec/Hz is assumed for successful decoding at $D$. 
If the direct transmission rate exceeds $R_{th}$, 
$S$ transmits a new symbol in the second phase. 
Otherwise, $R$ forwards the decoded symbol to $D$ 
only if the $SR$ link can guarantee a certain rate in the second phase. 
It is assumed that the total power, $P_T$, is divided equally among $S$ and $R$.

\subsection{Channel Model}
\label{channel_model}
The received symbol $y_{i}$ through the $i$th link is expressed as
\beqarr
y_{i}= \sqrt{P_{R_{i}}} h_{i} s+z_{i}\,, \; \; i\in \{SD, SR, RD\},
\label{eq_revd}
\eeqarr
where $P_{R_{i}}$ is the received power, $h_{i}$ is the channel gain of the $i$th link, $z_{i}$ is the additive noise sample at the receiver, and $s$ is the unit power transmitted symbol. 
The received power $P_{R_i}$ depends on the transmit power, length of the power cable, and path loss.
The channel gain multiplier $h_i$ is modeled as an independently distributed log-normal r.v. with
PDF
\beqarr
f_{h_{i}}(v) = \frac{1}{v\sqrt{2\pi \sigma_{h_{i}}^2}}
\exp\left(-\frac{1}{2}\left(
\frac{\ln{v} - \mu_{h_{i}}}{\sigma_{h_{i}}}
\right)^2
\right) \, , \; \; v \geq 0,
\label{e12}
\eeqarr
where the parameters $\mu_{h_{i}}$ and $\sigma_{h_{i}}$ are the mean
and the standard deviation of the normal r.v. $\ln{(h_{i})}$, respectively.
The $\ell$th moment of $h_{i}$ is given by
\beqarr
\mathbb{E}\left[h_{i}^{\ell}\right] = \exp\left({\ell}\mu_{h_i} +
\frac{{\ell}^2\sigma_{h_i}^2}{2}\right) \,.
\label{e13}
\eeqarr
We assume unit energy of the channel gain, i.e., $\mathbb{E}[h_{i}^2]=1$. According to (\ref{e13}),
this implies $\mu_{h_i}=-\sigma_{h_i}^2 \,.$

The dB equivalent of the received power through the $i$th link, $i\in\{SD, SR, RD \}$, 
$P_{R_i}$, can be expressed as
\beqarr
P_{R_{i}} \mbox{(dB)}=P_{T_{S}} \mbox{(dB)}-d_{i}\mbox{(km)}\times P_L \mbox{(dB/km)}\,,
\label{e9}
\eeqarr
where $P_L\mbox{(dB/km)}$ denotes the distance-dependent path loss factor.

\subsection{SNR Distribution}
\label{snr_model}
The symbols transmitted through power lines
suffer from impulsive noise as well as background noise \cite{PLC_BOOK:10}.
We assume the Bernoulli-Gaussian model \cite{MaSoGu:05}
which is mostly used \cite{DuMaSc:15}.
Thus, the additive noise sample $z_{i}$
can be written as
\beqarr
z_{i} = z_{W_i} + z_{B_i} z_{I_i} \,,
\label{e15}
\eeqarr
where $z_{W_i}$ and $z_{I_i}$
represent the background and impulsive noise samples, respectively,
and $z_{B_i}$ is a Bernoulli r.v. which equals $1$ with probability $p$ and $0$ with probability $(1-p)$.
The samples $z_{W_i}$ and $z_{I_i}$
are taken from the Gaussian distribution with mean zero and variance
$\sigma_W^2$ and $\sigma_I^2$, respectively. As background and impulsive noises have different origin, $z_{W_{i}}$, $z_{I_i}$, and $z_{B_i}$ are independent \cite{GoRaDo:04}.
Therefore, the noise samples $z_i$ are independent and identically distributed (i.i.d.)
r.v.s, each with PDF \cite{MaSoGu:05}
\beqarr
p_{z_{i}}(\nu) =\sum\limits_{j=1}^{2} \frac{p_j}
{\sqrt{2\pi \sigma_{j}^2}}\exp{\left(\frac{-\nu^2}
{2\sigma_{j}^2}\right)}
\, ,
\label{e16}
\eeqarr
where
$p_1=1-p \, , \; \; p_2=p \, , \; \; \sigma_1^2=\sigma_W^2 \, ,
\; \;
\sigma_2^2=\sigma_W^2+\sigma_I^2 \,$ .
The average noise power, $N_{0_i}=\mathbb{E}\left[ z_{i}^2 \right]$,
is given as
\beqarr
&N_{0_i}
    = \mathbb{E}\left[ z_{W_i}^{2} \right]+\mathbb{E}\left[
    z_{B_i}^2 \right]\mathbb{E}\left[ z_{I_i}^{2} \right]
    = \sigma_W^2 (1+p\ \eta)\, ,
\label{e18}
\eeqarr
where
$\eta=\frac{\sigma_I^2}{\sigma_W^2}$
represents the power ratio of impulsive noise to background noise.

As the channel gain $h_{i}$ is log-normally distributed,
the corresponding instantaneous SNR, 
$\gamma_{i}=\frac{{P}_{R_{i}}h_{i}^2}{N_{0_i}}$,
is also log-normally distributed with PDF
\beqarr
f_{\gamma_{i}}(w) =
\frac{1}{w\sqrt{2\pi \sigma_{\gamma_i}^2}}
\exp\left(-\frac{1}{2}
\left(\frac{\ln {w} - \mu_{\gamma_i}}
{\sigma_{\gamma_i}}
\right)^2
\right) \, , \; \; w \geq 0,
\label{e21}
\eeqarr
and parameters
$\mu_{\gamma_i} = 2\mu_{h_i}+\ln{\frac{{P}_{R_i}}{N_{0_i}}}\,,
\quad \sigma_{\gamma_i} = 2\sigma_{h_i}\,$.
The CDF of $\gamma_i$ is therefore given by
\beqarr
F_{\gamma_{i}}(w)=
\Pr[\gamma_{i}\leq w]=
1-Q\left(
\frac{\ln{w}-\mu_{\gamma_i}}{\sigma_{\gamma_i}}
\right) \, , \; \; w \geq 0,
\label{e23}
\eeqarr
where $Q(\cdot)$ denotes the Gaussian $Q$-function.

\subsection{Required SNR Threshold}
\label{snr_threshold}
As the channel is corrupted by background noise with probability 
$p_1=(1-p)$, and background and impulsive noise with probability $p_2=p$, the 
instantaneous channel capacity can be expressed as \cite{WiStTu:09} 
\beqarr
\label{eq_capacity}
C_i(\gamma_i)=
\sum_{j=1}^{2}
p_j\log_2{\left(1+\alpha_j\gamma_i\right)}\,\,,
\label{e24}
\eeqarr
where 
$\alpha_1= \frac{1+p\eta}{2}$ 
and 
$\alpha_2= \frac{1+p\eta}{2(1+\eta)}$  \cite{ShMaMiDuRa:16}.
Therefore, for the successful detection of the signal from the direct link, the approximate SNR threshold
that should be maintained at $D$ corresponding to the rate requirement $R_{th}$ can be obtained from (\ref{eq_capacity}) as
\beqarr
\Gamma_{SD}\approx {\alpha_1}^{-p1}{\alpha_2}^{-p2}2^{R_{th}}\,.\quad
\label{e26}
\eeqarr
To maintain the same rate requirement at $D$ through the half-duplex relayed path, the
$SR$ or $RD$ link should maintain twice the rate of the $SD$ link, and hence, the required SNR
threshold is
\beqarr
\Gamma_{SR}=\Gamma_{RD}\approx {\alpha_1}^{-p1}{\alpha_2}^{-p2}2^{2R_{th}}\,.\quad
\label{e26_1}
\eeqarr

\section{Outage Probability}
\label{sec_outage}
The outage probability is defined as the probability that the instantaneous channel capacity falls
below a predefined rate. An outage event would occur if any of the following events happens: i) the transmitted symbol cannot be detected both from the $SD$ and $SR$ links, or ii) the $SD$ link fails to detect the symbol, and, even if $R$ is able to correctly forward it, $RD$ link fails to deliver. Thus, the outage probability can be expressed mathematically by summing up the events i) and ii) as
\begin{align}
\hspace{-0.75cm}
{\cal P}_{o}(R_{th})
&={\Pr}\left[\gamma_{SD}<\Gamma_{SD}\right]
{\Pr}\left[\gamma_{SR}<\Gamma_{SR}\right]
\nn\\
&
+
{\Pr}\left[\gamma_{SD}<\Gamma_{SD}\right]
{\Pr}\left[\gamma_{SR}>\Gamma_{SR}\right]
\nn\\
&
\times
{\Pr}\left[\gamma_{RD}<\Gamma_{RD}\right].
%{\Pr}\left[\gamma_{RD}<{\alpha_1}^{-p1}{\alpha_2}^{-p2}2^{2R_{th}}\right].
\label{e36}
\end{align}
Finally, using (\ref{e23}) and after some algebra, the outage probability can be expressed 
in closed-form as
\begin{align}
{\cal P}_{o}(R_{th})&=
Q\left(
\frac{\mu_{\gamma_{SD}}-\ln\left(\Gamma_{SD}\right)}
{\sigma_{\gamma_{SD}}}
\right)
Q\left(
\frac{\mu_{\gamma_{SR}}-\ln\left(\Gamma_{SR}\right)}
{\sigma_{\gamma_{SR}}}
\right)
\nn\\
&+
Q\left(
\frac{\mu_{\gamma_{SD}}-\ln\left(\Gamma_{SD}\right)}
{\sigma_{\gamma_{SD}}}
\right)
Q\left(
\frac{\ln\left(\Gamma_{SR}\right)-\mu_{\gamma_{SR}}}
{\sigma_{\gamma_{SR}}}
\right)\nn\\
&
\times
Q\left(
\frac{\mu_{\gamma_{RD}}-\ln\left(\Gamma_{RD}\right)}
{\sigma_{\gamma_{RD}}}
\right).
%\times
%Q\left(
%\frac{\mu_{\gamma_{RD}}-\ln\left(\alpha_{1}^{-p1}
%\alpha_{2}^{-p2}
%2^{2 R_{th}}\right)}
%{\sigma_{\gamma_{RD}}}
%\right).
\label{e37}
\end{align}

\section{Relay Usage}
\label{sec_relay_usage}
%It is clear that t
The more the relay is used for data transmission,
the poorer the spectral efficiency, and the more the additional complexity
and delay required in data processing.
Hence, the fraction of times the relay is in use is of great interest and for the ISDF strategy. 
This number can be obtained by finding the probability that the $SD$ link fails whereas 
the $SR$ link attains the required rate threshold and is expressed as
\begin{align}
&N=
\mbox{Pr}[\gamma_{SD}<\Gamma_{SD}]\mbox{Pr}[\gamma_{SR}>\Gamma_{SR}]\nn\\
&= 
%\left(
Q\left(
\frac{\mu_{\gamma_{SD}}-\ln\left(\Gamma_{SD}\right)}
{\sigma_{\gamma_{SD}}}
\right)
%\right)\nn\\
%&\times&
Q\left(
\frac{\ln\left(\Gamma_{SR}\right)-\mu_{\gamma_{SR}}}
{\sigma_{\gamma_{SR}}}
\right).
\label{eq_avguse2}
%\eeqarr
\end{align}

\section{Average BER}
\label{sec_aber}
A bit error can occur either in the direct transmission or in the relayed transmission to $D$, according to 
the selective relaying technique assumed. A bit error in the direct transmission can occur in two ways: i) if its SNR exceeds the required threshold, or, ii) if its SNR does not exceed the required threshold and the relayed transmission is not used. Now, a bit error in the relayed transmission can occur if only one of the links between $SR$ or $RD$ is in error when the $SR$ link SNR exceeds the required threshold. 
%Mathematically, the average BER can be written by summing up all the above events for binary signaling scheme as
The average BER for binary signaling can be written by summing up the
probabilities of all the above events as
\begin{align}
P_{e}
&=
{\mathbb E}\left[P_e(\gamma_{SD}|\gamma_{SD}\geq\Gamma_{SD})\right]\nn\\
&+\Pr\left[\gamma_{SR}<\Gamma_{SR}\right]
{\mathbb E}\left[P_e(\gamma_{SD}|\gamma_{SD}<\Gamma_{SD})\right]\nn\\
&+\Pr\left[\gamma_{SD}<\Gamma_{SD}\right]
\left(1-{\mathbb E}\left[P_e(\gamma_{SR}|\gamma_{SR}\geq\Gamma_{SR})\right]\right)
\nn\\
&\times  {\mathbb E}\left[P_e(\gamma_{RD}|\gamma_{SR}\geq\Gamma_{SR})\right]
\nn\\
&+\Pr\left[\gamma_{SD}<\Gamma_{SD}\right]{\mathbb E}\left[P_e(\gamma_{SR}|\gamma_{SR}\geq\Gamma_{SR})\right]
\nn\\
&\times \left(1-{\mathbb E}\left[P_e(\gamma_{RD}|\gamma_{SR}\geq\Gamma_{SR})\right]\right).
\label{e27}
\end{align}

%For equiprobable binary phase shift-keying (BPSK) signaling scheme, the instantaneous BER 
For the equiprobable BPSK signaling scheme, the instantaneous BER 
as a function of $\gamma$, $P_e(\gamma)$\footnote{The subscript $i\in\{SR, RD, SD\}$ is dropped here 
onward to explain the relationship between BER and SNR in general.},  is given 
%similar to (\ref{eq_capacity})
as 
\beqarr
P_e(\gamma)=\sum_{j=1}^{2}p_jQ\left(\sqrt{\alpha_j\,\gamma} \right).
\label{e28}
\eeqarr
Thus, to obtain a closed-form expression for the average BER in (\ref{e27}),
we need the expectation operation of an integral of the type
\beqarr
{\mathbb E}[P_e(\gamma|y_1<\gamma\leq y_2)]=
\sum_{j=1}^{2}\int_{y_1}^{y_2}
p_j\,Q\left(\sqrt{\alpha_{j} y}\right)f_{\gamma}(y){\mbox{d}}y.
\label{e29}
\eeqarr
As $\gamma$ follows the log-normal distribution as given in (\ref{e21}),
we can write (\ref{e29}) as
\begin{align}
&{\mathbb E}[P_e(\gamma|y_1<\gamma\leq y_2)]
% \hspace{-2.25cm}
%\nn\\&
=\sum_{j=1}^{2}\int_{y_1}^{y_2}
p_j\,Q\left(\sqrt{\alpha_{j} y}\right)\nn\\
&\times
\frac{1}{y\sqrt{2\pi\sigma_{\gamma}^2}}
\exp{\left(-\frac{1}{2}\left(
\frac{\ln{(y)}-\mu_{\gamma}}{\sigma_{\gamma}}\right)^2
\right)}{\mbox{d}}y.
\label{e30}
\end{align}
Using the transformation $\ln(y)=2t-\ln(\alpha_{j})\label{e31}$,
(\ref{e30}) can be rewritten as
\begin{align}
\hspace{-0.45cm}
&{\mathbb E}[P_e(\gamma|y_1<\gamma\leq y_2)]
% \hspace{-2cm}
%\nn\\&
=
\sum_{j=1}^{2}\int_{\ln(\sqrt{\alpha_{j}y_1})}^{\ln(\sqrt{\alpha_{j}y_2})}
p_j\,Q\left(\exp(t)\right)\nn\\
&\times
\frac{2}{\sqrt{2\pi\sigma_{\gamma}^2}}
\exp{\left(-\frac{1}{2}\left(
\frac{2t-\ln(\alpha_j)-\mu_{\gamma}}{\sigma_{\gamma}}\right)^2
\right)}{\mbox{d}}t.
\label{e32}
\end{align}

It is difficult to evaluate the above integral in closed-form, as it
contains a function of the form $Q(\exp(t))$.
Therefore, we propose a novel approximation using the curve fitting technique to deal with such a function.
The approximation is given as
\beqarr
Q(\exp(t))\approx\sum_{m=1}^{M}
a_m \exp\left(-\left(\frac{t-b_m}{c_m}\right)^2\right)\,,
\label{e33}
\eeqarr
where $a_m, b_m,$ and $c_m$ are fitting constants. The number of summation terms, $M$,
depends on the region of interest and accuracy of the fit. A suitable value of $M$ and
corresponding $a_m, b_m,$ and $c_m$ values are further discussed
in Section \ref{sec_results}. Using the approximation in (\ref{e33}),
the integral in (\ref{e32}) can be evaluated in approximate closed-form as
\begin{align}
%\hspace{-1cm}
&{\mathbb E}[P_e(\gamma|y_1<\gamma\leq y_2)]\nn\\
% \hspace{-3cm}
&\approx \sum_{m=1}^{M}
\sum_{j=1}^{2}\int_{\ln(\sqrt{\alpha_{j}y_1})}^{\ln(\sqrt{\alpha_{j}y_2})}
p_j\,a_m \exp\left(-\left(\frac{t-b_m}{c_m}\right)^2\right)
\nn\\
&\times
\frac{2}{\sqrt{2\pi\sigma_{\gamma}^2}}
\exp{\left(-\frac{1}{2}\left(
\frac{2t-\ln(\alpha_j)-\mu_{\gamma}}{\sigma_{\gamma}}\right)^2
\right)}{\mbox{d}}t\,
%\nn\\
%&=&
%\sum_{m=1}^{M}
%\sum_{j=1}^{2}\int_{\ln(\sqrt{\alpha_{j}y_1})}^{\ln(\sqrt{\alpha_{j}y_2})}
%p_j\,a_m \nn\\
%&&
%\times
%\frac{2}{\sqrt{2\pi\sigma_{\gamma}^2}}
%\exp\left(-\left(A_{m}^2t^2-2B_{m,j}t+C_{m,j}\right)\right)
%{\mbox{d}}t
\nn\\
&=
\sum_{m=1}^{M}
\sum_{j=1}^{2}
\frac{2\,p_j\,a_m}{\sigma_{\gamma}\sqrt{2}A_{m}}
\exp\left(
-\left(
C_{m,j}-\left(\frac{B_{m,j}}{A_{m}}\right)^2
\right)
\right)
 \nn\\
 &
 \times
\left\{
Q\left(\sqrt{2}\left(A_{m}\ln(\sqrt{\alpha_{j}y_1})-\frac{B_{m,j}}{A_{m}}\right)\right)
\right.
\nn\\
&
\left.
-
Q\left(\sqrt{2}\left(A_{m}\ln(\sqrt{\alpha_{j}y_2})-\frac{B_{m,j}}{A_{m}}\right)\right)
\right\}\,,
\label{e34}
\end{align}
where
%\begin{align}
$A_{m}=\sqrt{\frac{1}{c_m^2}+
\frac{2}{\sigma_{\gamma}^2}}$,
%\,,
%\nn\\
%\,\,
$B_{m,j}=\frac{b_m}{c_m^2}+
\frac{\ln(\alpha_j)+\mu_{\gamma}}{\sigma_{\gamma}^2}$, 
%\,,\nn\\
$C_{m,j}=\frac{b_m^2}{c_m^2}+
\frac{\left(\ln(\alpha_j)+\mu_{\gamma}\right)^2}{2\sigma_{\gamma}^2}$.
%\,.
%\label{e35}
%\end{align}
Finally, using (\ref{e34}), the average BER in (\ref{e27}) 
can be expressed in approximate closed-form as in (\ref{e35a}).

\begin{table*}
\begin{tabular}{m{\textwidth}}
%\hline
{
% \begin{align}
\beqarr
% \hspace{-0.5cm}
&P_{e}
% \hspace{-1cm}
% &&
% \nn\\
&=
\sum_{m=1}^{M}
\sum_{j=1}^{2}
\frac{2\,p_j\,a_m}{\sigma_{\gamma_{SD}}\sqrt{2}A_{m,{SD}}}
% \nn\\
% &&\times
\exp\left(
-\left(
C_{m,j,{SD}}-\left(\frac{B_{m,j,{SD}}}{A_{m,SD}}\right)^2
\right)
\right)
%  \nn\\
% &&
\left[
Q\left(\sqrt{2}\left(A_{m,SD}\ln(\sqrt{\alpha_{j}\Gamma_{SD}})-\frac{B_{m,j,{SD}}}{A_{m,{SD}}}\right)\right)
\right]
\nn\\
&&
+
\left(
1-Q\left(
\frac{\ln\Gamma_{SR}-\mu_{\gamma_{SR}}}{\sigma_{\gamma_{SR}}}
\right)
\right)
% \nn\\
% &&
% \times
\sum_{m=1}^{M}
\sum_{j=1}^{2}
\frac{2\,p_j\,a_m}{\sigma_{\gamma_{SD}}\sqrt{2}A_{m,{SD}}}
% \nn\\
% &&
% \times
\exp\left(
-\left(
C_{m,j,{SD}}-\left(\frac{B_{m,j,{SD}}}{A_{m,SD}}\right)^2
\right)
\right)
 \nn\\
&&
\times
\left[1-
Q\left(\sqrt{2}\left(A_{m,SD}\ln(\sqrt{\alpha_{j}\Gamma_{SD}})-\frac{B_{m,j,{SD}}}{A_{m,{SD}}}\right)\right)
\right]
% \nn\\
% &&
+
\left(
1-Q\left(
\frac{\ln\Gamma_{SD}-\mu_{\gamma_{SD}}}{\sigma_{\gamma_{SD}}}
\right)
\right)
\nn\\
&&
\times
\left\lbrace
\Bigg(
1-
\sum_{m=1}^{M}
\sum_{j=1}^{2}
\frac{2\,p_j\,a_m}{\sigma_{\gamma_{SR}}\sqrt{2}A_{m,{SR}}}
\right.
% \nn
% \\
% &&
\left.
\exp\left(
-\left(
C_{m,j,{SR}}-\left(\frac{B_{m,j,{SR}}}{A_{m,SR}}\right)^2
\right)
\right)
\right.
\Bigg)
% \nn\\
% &&
% \times
\Bigg(
\sum_{m=1}^{M}
\sum_{j=1}^{2}
\frac{2\,p_j\,a_m}{\sigma_{\gamma_{RD}}\sqrt{2}A_{m,{RD}}}
\nn
\\
&&
\times\left.
\exp\left(
-\left(
C_{m,j,{RD}}-\left(\frac{B_{m,j,{RD}}}{A_{m,RD}}\right)^2
\right)
\right)
\right.
% \nn\\
% &&
% \times
\left(
Q\left(
\frac{\ln\Gamma_{SR}-\mu_{\gamma_{SR}}}{\sigma_{\gamma_{SR}}}
\right)
\right)\Bigg)\nn
% \\
% &&
+
\Bigg(
\sum_{m=1}^{M}
\sum_{j=1}^{2}
\frac{2\,p_j\,a_m}{\sigma_{\gamma_{SR}}\sqrt{2}A_{m,{SR}}}
\nn
\\
&&
\times
\left.
\exp\left(
-\left(
C_{m,j,{SR}}-\left(\frac{B_{m,j,{SR}}}{A_{m,SR}}\right)^2
\right)
\right)
\right.
\Bigg)
% \nn\\
% &&
% \times
\Bigg(
1-
\sum_{m=1}^{M}
\sum_{j=1}^{2}
\frac{2\,p_j\,a_m}{\sigma_{\gamma_{RD}}\sqrt{2}A_{m,{RD}}}
% \nn
% \\
% &&
\left.
\exp\left(
-\left(
C_{m,j,{RD}}-\left(\frac{B_{m,j,{RD}}}{A_{m,RD}}\right)^2
\right)
\right)
\right.
\nn\\
&&
\times
\left(
Q\left(
\frac{\ln\Gamma_{SR}-\mu_{\gamma_{SR}}}{\sigma_{\gamma_{SR}}}
\right)
\right)
\Bigg)\Bigg\}.
\label{e35a}
\eeqarr
%%%%%%%%%%%%%%%%%%%%%%%%
% \end{align}
}
\\
\hline
\end{tabular}
\end{table*}

\begin{figure}
%[t]
\psfig{file=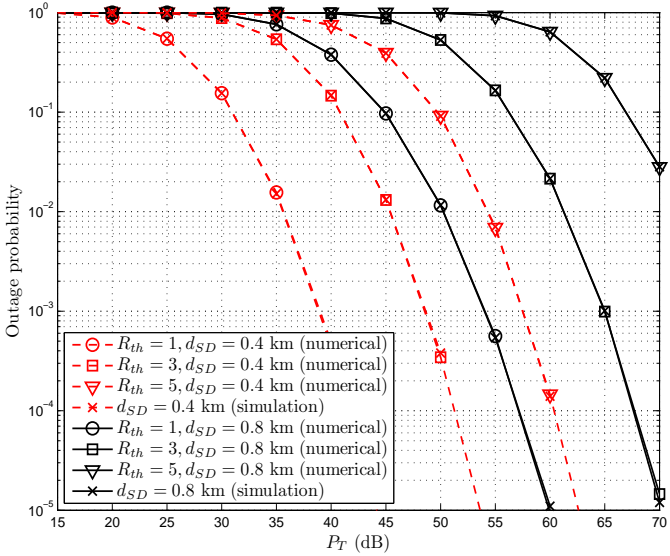,width=9cm,height=7.5cm}
\caption{Outage probability versus total transmit power with
$\sigma_{h_{i}} =3$ dB, ${P}_{L}=60$ dB/km, $p = 0.1$, and  $\eta = 10$
for different values of $R_{th}$ and $d_{SD}$.}
\label{fig_5}
\end{figure}

\begin{figure}
%[t]
\psfig{file=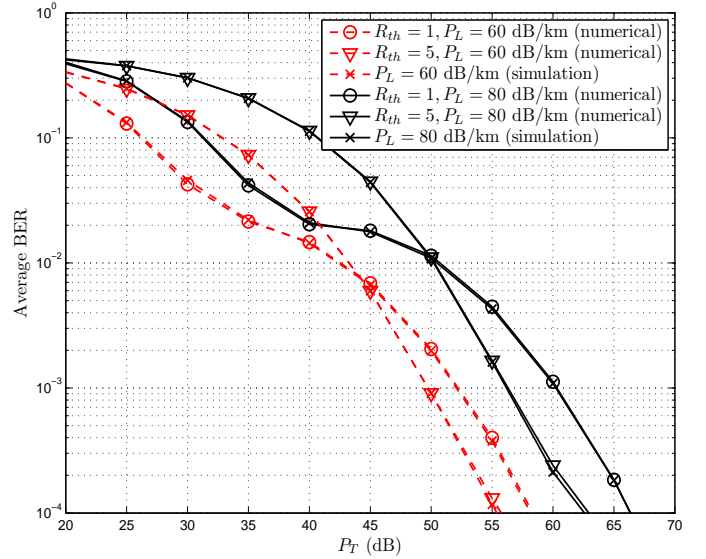,width=9cm,height=7.5cm}
\caption{Average BER versus total transmit power with
$\sigma_{h_i} =3$ dB, $d_{SD}=0.4$ km, $p = 0.1$, and  $\eta = 10$
for different values of $R_{th}$ and ${P}_{L}$.}
\label{fig_3}
\end{figure}

\section{Results and Discussions}
\label{sec_results}

\begin{table}
% [h]
\caption{Parameters in (\ref{e33}) from the curve fitting for $M=7$.}
\begin{center}
\begin{tabular}{| c | c | c|c|}
 \hline $m$ & $a_{m}$ & $b_{m}$ & $c_{m}$\\
   \hline
$1$ & 0.4665 & -5.37 & 2.174\\
\hline
$2$ & -0.0007029 & -3.674 & 0.1178\\
\hline
$3$ & 0.0165 & -3.141 & 0.0004957 \\
\hline
$4$ & 0.2831 & -2.998 & 1.458\\
\hline
$5$ & 0.2113 & -1.764 & 1.06\\
\hline
$6$ & 0.1742 & -0.8425 & 0.837\\
\hline
$7$ & 0.07986 & -0.1109 & 0.6399\\
\hline
\end{tabular}
\end{center}
\label{tab1}
\end{table}

Numerical and simulation results are presented here to validate the performance analysis. 
Unless otherwise mentioned, the following parameters are considered. 
$SD$ is chosen as $400$ m and $800$ m, respectively, in consistence with a small PLC system environment
\cite{ShMaMiDuRa:16}. 
Depending on the power distribution network, in general, $\sigma_{h_{i}}$ lies in between
$2$ dB to $5$ dB \cite{GuCeAr:11}. Here we assume $\sigma_{h_{i}}=3$ dB, $\forall i$, where the conversion from absolute scale to dB scale is given by $\sigma_{h_{i}} (\text{dB}) = 10\sigma_{h_{i}}/ \ln{10}$. A high value of $\sigma_{h_{i}}$ indicates high fluctuation in the received signal power \cite{PaCaKaTh:03,GuCeAr:11}. The distance-dependent path loss factor depends upon the type of cable and carrier frequency used for the transmission, and ranges from $40$ to $100$ dB/km \cite{Ho:98}. Hence, $P_L=60$ and $80$ dB/km are chosen, respectively.
The values of the impulsive noise parameters are $p=0.1$ and $\eta=10$, following \cite{DuMa:16}. Fitting constants for the approximation in (\ref{e33}) are obtained from the curve fitting tool of MATLAB with $M=7$, root mean squared error (RMSE) $0.0006931$, and sum of squares due to error (SSE) of $0.0004708$. The parameters calculated from the curve fitting are given in Table \ref{tab1}.

Fig. \ref{fig_5} shows the outage probability versus ${P}_{T}$, for different values of
$R_{th}$ and $d_{SD}$. The numerical curves are obtained using (\ref{e37}) and are found to agree well with the simulation results, thus validating our outage analysis.
% From Fig. \ref{fig_5}, t
To achieve an outage probability of $10^{-3}$ with $R_{th}=3$ bits/sec/Hz, the PLC system with $d_{SD}=0.4$ km requires $P_{T}=48$ dB, whereas with $d_{SD}=0.8$, the transmit power requirement increases to $65$ dB. Thus, it can be concluded that for a fixed $R_{th}$ and $P_T$, the outage performance degrades with increasing $d_{SD}$.

Fig. \ref{fig_3} shows the average BER versus ${P}_{T}$ for different $R_{th}$ and $P_L$ values.
The numerical curves are obtained using the approximate closed-form expression derived in (\ref{e35a}). 
The numerical results are also in agreement with the simulation results, thus validating our analysis. In general, it is observed that the performance improves as $P_T$ increases and also for fixed $P_T$ and $R_{th}$, the performance degrades with increasing the distance-dependent path loss. 
%Further, it is noticed that when $P_T$ is low, the performances degrades with increasing $R_{th}$; 
%however, when $P_T$ is high, the performance improves with increasing $R_{th}$. When $R_{th}$ increases, generally, the average BER decreases at a lower $P_T$ as neither $SD$ nor 
%$SR$ can overcome the increased SNR threshold at $D$ and $R$, respectively.
%If $P_T$ is increased further, $R$ can eventually overcome the required SNR threshold due to comparatively low path loss and increased received power at it, hence, the observation. 
Further, it is noticed that when $P_T$ is low, the performances degrades with increasing $R_{th}$; 
however, when $P_T$ is high, the performance improves with the increase in $R_{th}$.
This is an interesting observation as with the increase in $R_{th}$, intuitively, the average BER should degrade
at all SNRs. When $R_{th}$ increases, the average BER decreases at a lower $P_T$ as neither $SD$ nor 
$SR$ can overcome the increased SNR threshold at $D$ and $R$, respectively.
If $P_T$ is increased further, $R$ can eventually overcome the required SNR threshold due to comparatively low path loss and increased received power at it, hence, this observation. 
Moreover, at higher values of $P_T$ the strategy tends to follow the direct transmission, and hence, the BER curves for various $R_{th}$ are parallel. 

\begin{figure}
% [t]
\psfig{file=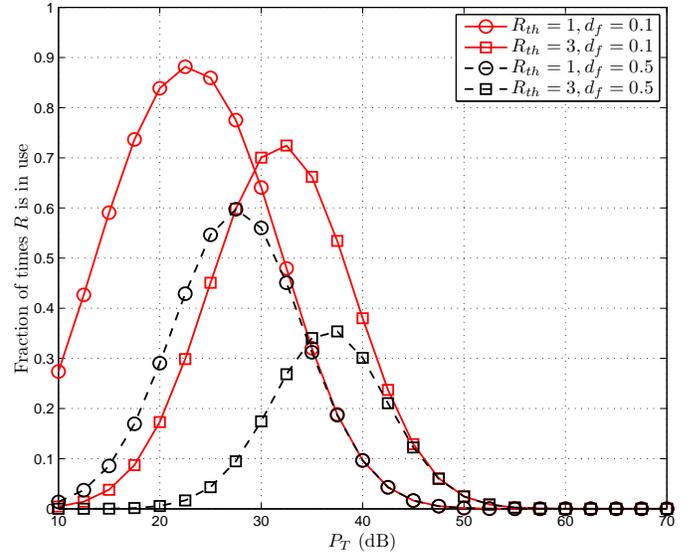,width=9cm,height=7.5cm}
\caption{Fraction of times $R$ is in use versus $P_T$ with
$\sigma_{h_i} =3$ dB, $d_{SD}=0.4$ km, ${P}_{L}=60$ dB/km, $p = 0.1$, and  $\eta = 10$
for different values of $R_{th}$ and $d_{f}$.}
\label{fig_4}
\end{figure}

In Fig. \ref{fig_4}, the fraction of times $R$ is in use for transmission versus $P_T$
is plotted using (\ref{eq_avguse2}) for different $R_{th}$ values and relay placements ($d_f$),   
where $d_{SR}=d_f d_{SD}$ and $0<d_f<1$. 
It is observed that for various $R_{th}$ and $d_f$, the curves are bell-shaped and never reach unity. As $P_T$ increases, initially relay usage increases due to improved $SR$ link quality, later relay usage decreases due to better direct link quality, and hence, the bell-shape.
Thus, it can be concluded that the ISDF is spectrally efficient when compared to the traditional DF relaying, which uses the relay in each transmission. 
Next, it is observed that as $d_f$ increases at a given $R_{th}$, the curves shift towards right and the maximum fraction of times the relay is in use also decreases. This can be explained by the fact that as the length of $SR$ link increases, the received SNR at the relay decreases, which in turn reduces the fraction of times the relay is in use.
%It can be seen from the combined analysis of Figs. {\ref{fig_5}} and {\ref{fig_4}} that to achieve an outage probability of $10^{-3}$ with $d_{SD}=0.4$ km, $d_f=0.5$, and $R_{th}=1$ bit/sec/Hz, the required $P_T$ is approximately $\approx 39$ dB and the fraction of times the relay is in use at this $P_T$ is approximately $10\%$.
%Thus, it can be concluded that ISDF is spectrally efficient when compared to the traditional DF relaying, which uses the relay in each transmission.  
Further, we can observe that at a given $d_f$ and beyond a certain $P_T$, the fraction of times the relay is in use for higher $R_{th}=3$ becomes more than lower $R_{th}=1$ due to the bell-shape. 
This means that the spectral efficiency decreases when $R_{th}$ increases at higher $P_T$.
This also justifies the crossovers of the average BER plots for the same $R_{th}$ beyond certain $P_T$ in Fig. \ref{fig_3}. 
Thus, although the spectral efficiency decreases at higher $P_T$ when $R_{th}$ increases, interestingly the average BER improves.

\section{Conclusion}
\label{sec_conclusion}
In this work, the 
%incremental selective decode-and-forward
ISDF relaying strategy has been introduced for PLC systems
to improve spectral efficiency. Closed-form expressions for the outage probability and
the fraction of times the relay is in use along with an approximate closed-form expression for the average BER are derived considering the BPSK signaling scheme. Log-normal fading and Bernoulli-Gaussian impulsive noise are considered for the analysis. It is observed that at lower transmit power, 
the performance degrades as the required rate, path loss, or end-to-end distance increases. 
It is found that the proposed relaying strategy can provide an overall improved 
spectral efficiency. Furthermore, although the spectral efficiency decreases at higher transmit power when the required rate increases, the average BER improves.

\section*{Acknowledgment}
\label{sec_acknowledgment}
This work was supported in part by the Department of Science and Technology (DST), Govt. of India (Ref. No. TMD/CERI/BEE/2016/059(G)), Royal Society-SERB Newton International Fellowship under Grant NF151345, and Natural Science and Engineering Council of Canada (NSERC), through its Discovery program.

\end{document}